\documentclass{emulateapj}

\usepackage{apjfonts}
\usepackage{amsmath,graphicx,lscape,natbib}
\slugcomment{Received 2011 April 17; accepted 2011 June 10; to appear in ApJL}

\newcommand{\angstrom}{{\rm \AA}}

\newcommand{\hbeta}{H{$\beta$}}
\newcommand{\halpha}{H{$\alpha$}}

\newcommand{\OI}{[O{\sevenrm\,I}]\,$\lambda$6300}

\newcommand{\OIIIb}{[O{\sevenrm\,III}]\,$\lambda$5007}

\newcommand{\NIIb}{[N{\sevenrm\,II}]\,$\lambda$6584}

\newcommand{\SIIa}{[S{\sevenrm\,II}]\,$\lambda$6717}
\newcommand{\SIIb}{[S{\sevenrm\,II}]\,$\lambda$6731}
\newcommand{\SIIab}{[S{\sevenrm\,II}]\,$\lambda\lambda$6717,6731}

 \font\sevenrm=cmr7 scaled 1000

\begin{document}

\title{Cosmic Train Wreck by Massive Black Holes: \\
Discovery of a kpc-Scale Triple Active Galactic
Nucleus\altaffilmark{1}}

\shorttitle{Triple AGN}

\shortauthors{Liu, Shen, \& Strauss}
\author{Xin Liu\altaffilmark{2,3}, Yue Shen\altaffilmark{2},
and Michael A. Strauss\altaffilmark{4}}

\altaffiltext{1}{This paper includes data gathered with the
Apache Point Observatory 3.5-meter telescope, which is owned
and operated by the Astrophysical Research Consortium.}

\altaffiltext{2}{Harvard-Smithsonian Center for Astrophysics,
60 Garden St., Cambridge, MA 02138}

\altaffiltext{3}{Einstein Fellow}

\altaffiltext{4}{Princeton University Observatory, Princeton,
NJ 08544}

\begin{abstract}
Hierarchical galaxy mergers will lead to the formation of
binary and, in the case of a subsequent merger before a binary
coalesce, triple supermassive black holes (SMBHs), given that
most massive galaxies harbor SMBHs. A triple of SMBHs becomes
visible as a triple active galactic nucleus (AGN) when the BHs
accrete large amounts of gas at the same time. Here we report
the discovery of a kpc-scale triple AGN, SDSS J1027+1749 at
$z=0.066$, from our systematic search for hierarchical mergers
of AGNs. The galaxy contains three emission-line nuclei, two of
which are offset by 450 and 110 km s$^{-1}$ in velocity and by
2.4 and 3.0 kpc in projected separation from the central
nucleus. All three nuclei are classified as obscured AGNs based
on optical diagnostic emission line ratios, with black hole
mass estimates $M_{\bullet} \gtrsim 10^8 M_{\odot}$ from
stellar velocity dispersions measured in the associated stellar
components. Based on dynamical friction timescale estimates,
the three stellar components in SDSS J1027+1749 will merge in
$\sim40$ Myr, and their associated SMBHs may evolve into a
gravitationally interacting triple system in $\lesssim200$ Myr.
Our result sets a lower limit of $\sim 5\times10^{-5}$ for the
fraction of kpc-scale triples in optically selected AGNs at
$z\sim0.1$.
\end{abstract}


%%%%%%%%%%%%%%%%%%%%%%%%%%%%%%%%%%%%%%%%%%%%%%%%%%%
\section{Introduction}\label{sec:intro}

Galaxies are thought to be built up hierarchically via mergers
\citep{toomre72}.  Because most massive galaxies are believed
to harbor a central supermassive black hole
\citep[SMBH;][]{kormendy95}, galaxy mergers will result in the
formation of binary SMBHs through dynamical friction
\citep{begelman80,milosavljevic01,yu02}.  When a binary does
not coalesce before a subsequent merger with a third galaxy, a
system of three gravitationally interacting SMBHs is expected
to form \citep[e.g.,][]{valtonen96}.  Triple SMBHs may hold
important clues to our general understanding of massive galaxy
formation. For example, numerical simulations suggest that
triple SMBHs scour out cores in stellar bulges with mass
deficits one or two times of the total mass of the BHs and
sizes larger than those formed around binary SMBHs. This
process may be responsible for the large cores observed in some
massive elliptical galaxies such as M87 \citep{hoffman07}.
Triple SMBHs also provide a unique astrophysical laboratory to
study the chaotic dynamics of three-body interactions in
General Relativity \citep[e.g.,][]{blaes02,merritt06}.
Numerical simulations of the dynamics of triples suggest that
their encounter may lead to a merger of all three BHs,
formation of a highly eccentric binary, or ejection of three
free BHs \citep{iwasawa06,lousto08}. Triple SMBHs exhibit
phases of very high eccentricity in the inner binary, producing
intense bursts of gravitational radiation which will be within
the sensitivity range of forthcoming pulsar timing arrays and
the Laser Interferometer Space Antenna
\citep[e.g.,][]{amaro10}.

Despite their significant scientific merit and intense
theoretical interest, direct observational evidence for
gravitationally interacting triple SMBHs is still lacking,
however, because their typical separation (less than a few
parsecs) is too small to be resolved at cosmological distances.
In this Letter, we report the probable discovery of a kpc-scale
triple of SMBHs, at a separation large enough that the
components can be easily resolved using current facilities, yet
small enough for the system to be dynamically interesting.
Using a simple stellar dynamical friction argument, we estimate
that the system will form a bound SMBH triple in $\lesssim200$
Myr. A triple of SMBHs becomes visible as a triple active
galactic nucleus (AGN) when the BHs accrete large amounts of
gas at the same time -- a process which is thought to be common
in gas-rich mergers \citep[e.g.,][]{hernquist89}. While
kpc-scale triple AGNs are supposed to be present in recurrent
galaxy mergers, only one candidate is known\footnote{There is
one probable physical triple quasar known, QQ $1429-008$ at
$z\sim2.1$ \citep{djorgovski07}, but the projected separations
between the quasar components are much larger, $\sim30$--50
kpc.}. It is a triplet of emission-line nuclei (with two offset
nuclei at 5.1 and 8.4 kpc in projection from the primary
nucleus) in the minor merger NGC 3341 at $z \sim 0.03$
serendipitously discovered by \citet{barth08}.

\begin{figure}
  \centering
    \includegraphics[width=42mm]{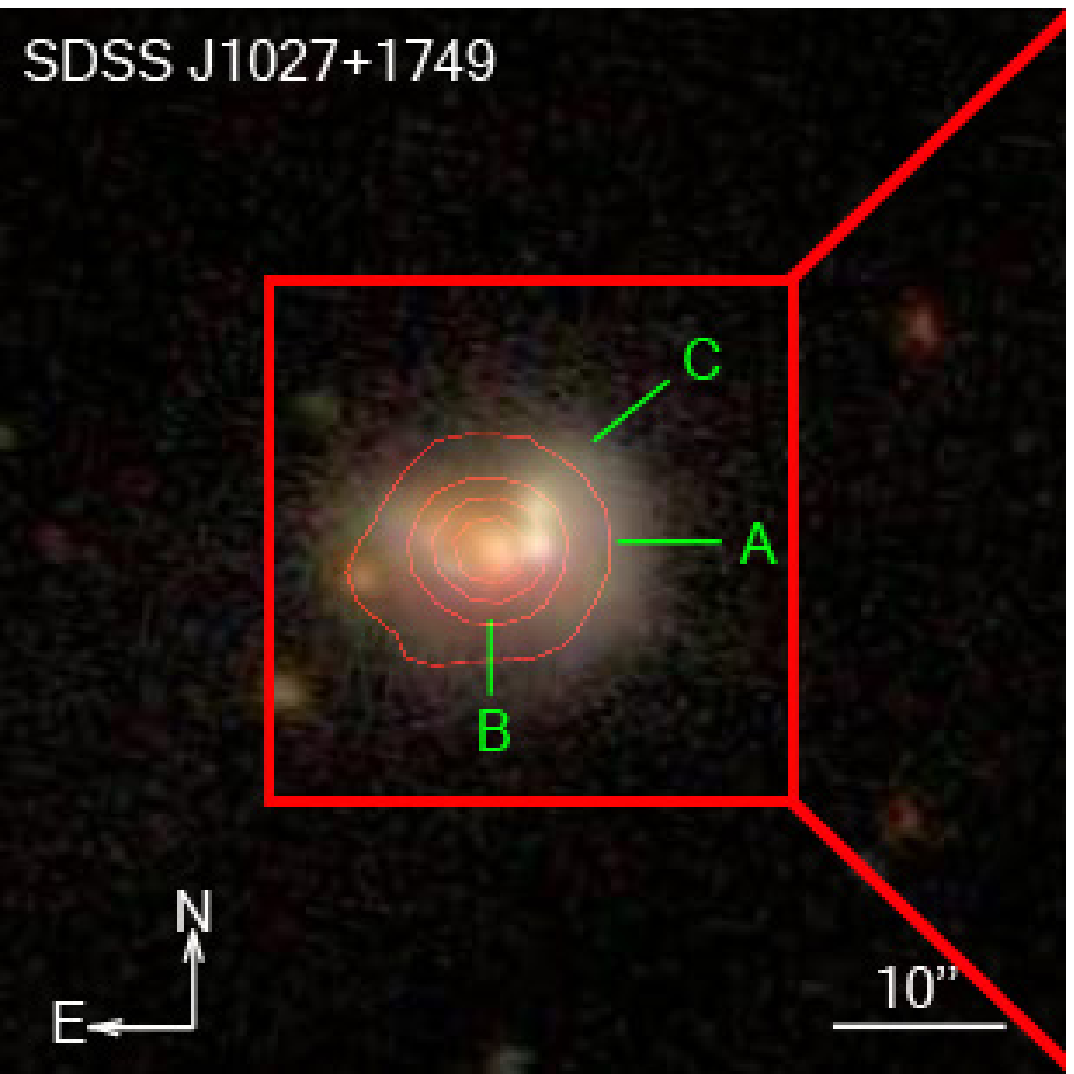}
    \includegraphics[width=42mm]{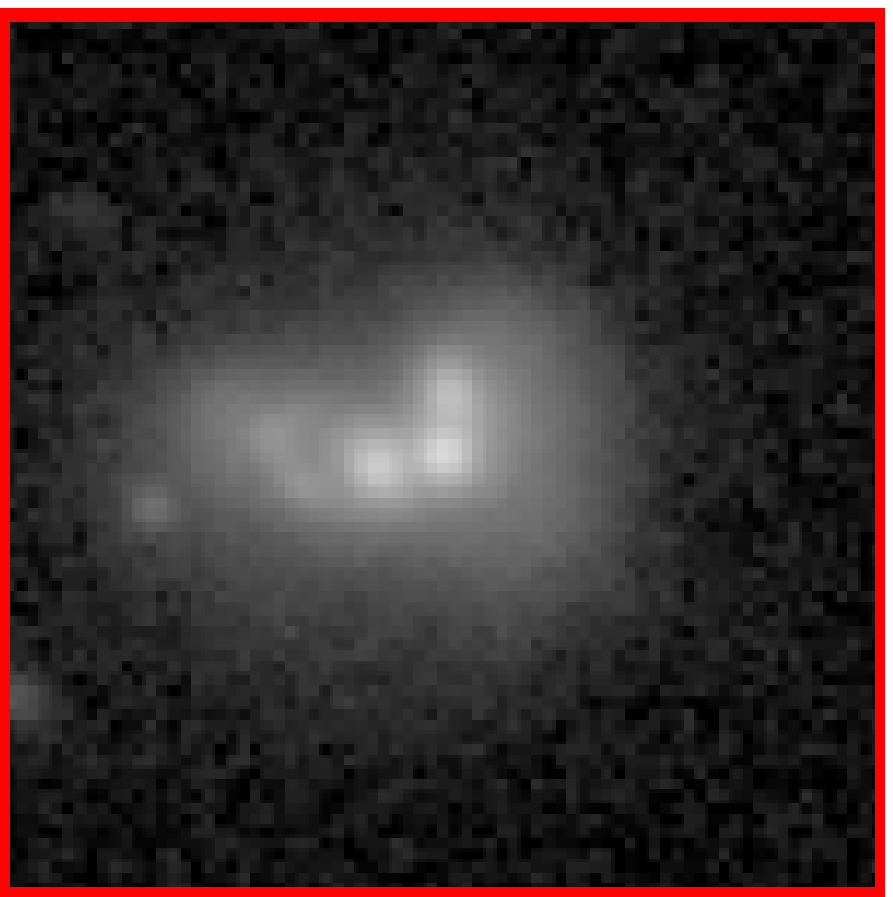}
    \caption{{\it Left}: SDSS $gri$-color composite image of SDSS J1027+1749,
    shown with a $1'\times1'$ FOV.
    Contours indicate radio flux densities from the FIRST 1.4 GHz survey
    \citep{becker95}.
    The galaxy has three nuclei A, B, and C, and a strong
    tidal feature to the northeast of B.
    Nucleus B (C) has a projected separation of 3.0 (2.4) kpc from A.
    {\it Right}: SDSS $r$-band image of the central $30''\times30''$ region.}
    \label{fig:sdssimg}
\end{figure}

To quantify the frequency of kpc-scale binary and triple
systems of SMBHs, we are conducting a systematic search for
hierarchical mergers of AGNs in the Seventh Data Release
\citep[DR7;][]{SDSSDR7} of the Sloan Digital Sky Survey
\citep[SDSS;][]{york00}. We have selected a sample of 1286 AGN
pairs with projected separations $r_p<100$ kpc and
line-of-sight velocity offsets $\Delta v < 600$ km s$^{-1}$
\citep{liu11a}. Ninety-two AGN pairs in the sample have $r_p <
10$ kpc and $z<0.16$.  Seven of these 92 pairs have a third
galaxy or nuclei with $r_p < 10$ kpc from both of the first two
nuclei and $r$-band magnitude differences $<4$ mag. Because of
its proximity with the first two nuclei, none of the third
nuclei has an SDSS spectrum due to the finite size of SDSS
fibers\footnote{The first two nuclei both have an SDSS spectrum
because they were observed on overlapping spectroscopic
plates.}. To determine whether the third nucleus is also active
and physically close to the first two nuclei, we are conducting
spatially resolved spectroscopy for these kpc-scale triple AGN
(hereafter, triple AGN, for short) candidates. Here, we report
our initial results on the discovery of a triple AGN in a
hierarchical merging system, SDSS J1027+1749. Shown in Figure
\ref{fig:sdssimg}, the galaxy contains three nuclei as seen in
its SDSS image, with B and C at projected separations of 3.0
and 2.4 kpc from A. The nuclei A and B have SDSS spectra. We
have conducted optical slit spectroscopy for all three nuclei.
We describe our observations and data analysis in \S
\ref{sec:obs}, and present physical measurements of the three
nuclei in \S \ref{sec:result}. We discuss implications of our
results and conclude in \S \ref{sec:discuss}. A $\Lambda$CDM
cosmology with $\Omega_m = 0.3$, $\Omega_{\Lambda} = 0.7$, and
$h = 0.7$ is assumed throughout.

%%%%%%%%%%%%%%%%%%%%%%%%%%%%%%%%%%%%%%%%%%%%%%%%%%%
\section{Observations and Data Analysis}\label{sec:obs}

We obtained slit spectra for SDSS J1027+1749 on the nights of
2011 March 2 and 6 UT using the Dual Imaging Spectrograph (DIS)
on the Apache Point Observatory 3.5 m telescope. The sky was
mostly non-photometric, with seeing ranging between $0.''8$ and
$2.''0$ with a median around $1.''2$. DIS has a 4$'\times$6$'$
FOV with a pixel size of $0.''414$. We adopted a
$1.''5\times6'$ slit with the B1200+R1200 gratings centered at
510 and 700 nm. The spectral coverage was 450--560 (640--760)
nm with an instrumental resolution of $\sigma_{{\rm inst}}
\approx 60$ (30) km s$^{-1}$ and a dispersion of 0.62 (0.58)
\angstrom\ pixel$^{-1}$ in the blue (red). We oriented the slit
with PA $= 105^{\circ}$, 133$^{\circ}$, and 172$^{\circ}$ to go
through the nuclei A and B, B and C, and A and C respectively,
as labeled on the SDSS image shown in Figure \ref{fig:sdssimg}.
The total effective exposure time was 10800, 8100, and 8100 s
for A, B, and C, respectively. Standard stars G191B2B and HZ44
were observed for spectrophotometric calibration. We reduced
the DIS data following standard IRAF\footnote{IRAF is
distributed by the National Optical Astronomy Observatory,
which is operated by the Association of Universities for
Research in Astronomy, Inc., under cooperative agreement with
the National Science Foundation.} procedures \citep{tody86}. We
extracted one-dimensional spectra using $3''$ diameter
apertures for the nuclei A, B, and C respectively. We applied a
telluric correction from standard stars on the extracted
one-dimensional spectra. The median signal-to-noise ratio (S/N)
achieved was 50--80 pixel$^{-1}$.

\begin{figure}
  \centering
    \includegraphics[width=87mm]{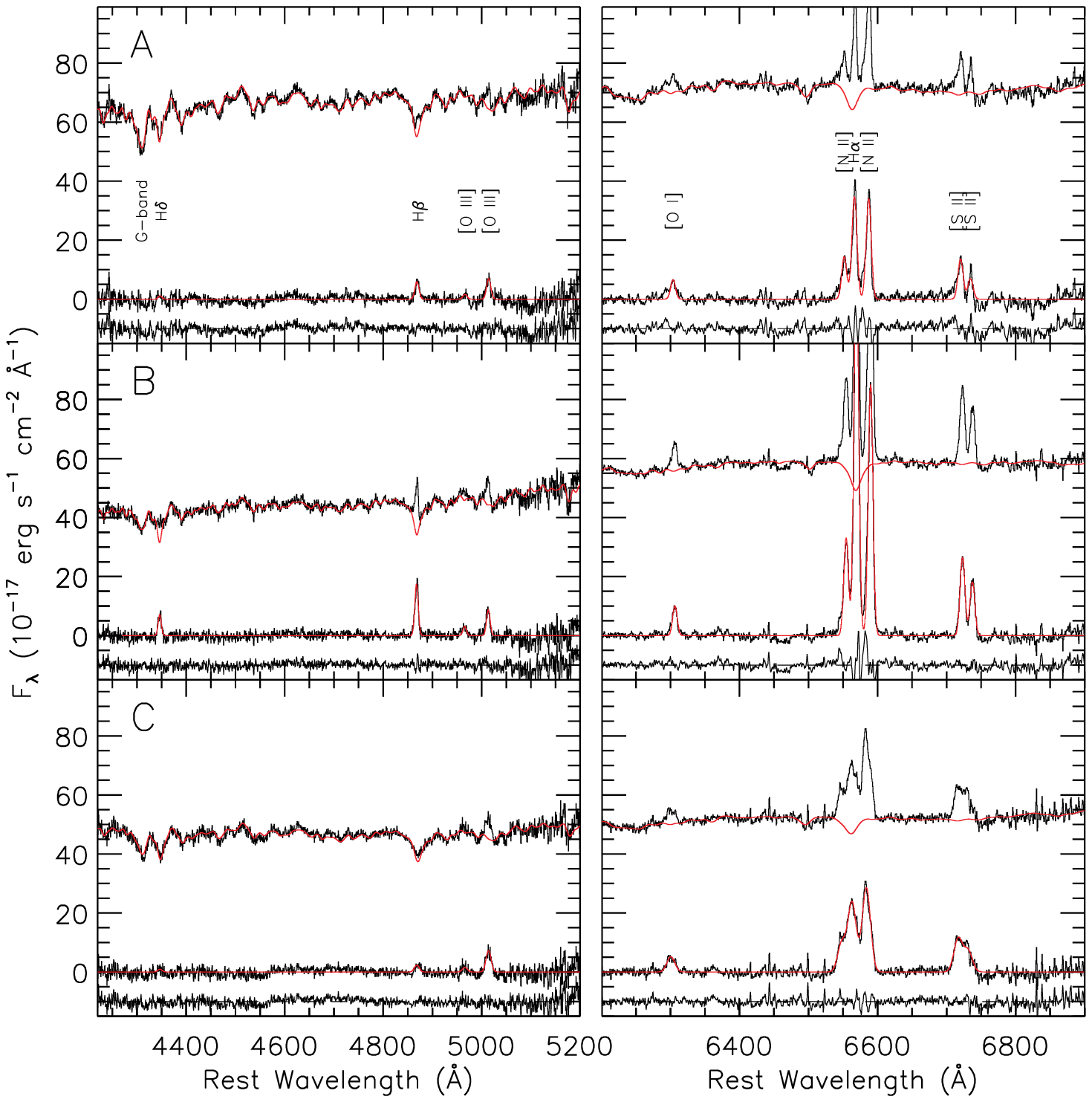}
    \caption{DIS one-dimensional spectra of the three nuclei of SDSS J1027+1749.
    We show the total spectra, the continuum-subtracted emission lines,
    and the continuum-and-emission-subtracted residuals offset
    by $-10^{-16}$ erg s$^{-1}$ cm$^{-2}$ \angstrom\ $^{-1}$.
    Plotted in red are our best-fit models for the continuum
    and emission lines, respectively.}
    \label{fig:dis1d}
\end{figure}

Figure \ref{fig:dis1d} shows the resulting rest-frame
one-dimensional spectra and our best-fit models for the stellar
continuum and the continuum-subtracted emission lines,
respectively. We measure redshift by fitting the continuum with
galaxy templates produced by population synthesis models of
\citet{bc03} using the procedure described in detail in
\citet{liu09}. The templates have been convolved with the
stellar velocity dispersions $\sigma_{\ast}$ measured over the
spectral region of 4130--4600 \angstrom\ containing the G band
$\lambda$4304 \angstrom\ using the direct fitting algorithm of
\citet{greene06a} and \citet{ho09} following the procedure
described in \citet{liu09}. After subtracting the stellar
continuum, we fit the emission lines simultaneously with
Gaussian models constrained to have the same redshift and
width.

%%%%%%%%%%%%%%%%%%%%%%%%%%%%%%%%%%%%%%%%%%%%%%%%%%%
\section{Results}\label{sec:result}

%\begin{landscape}
\begin{deluxetable*}{cccccccccccc}
\tabletypesize{\footnotesize}
%\tabletypesize{\scriptsize}
\tablecolumns{12} \tablewidth{\textwidth}
%\tablewidth{0pc}
%
\tablecaption{Host-galaxy Measurements of the Three Nuclei of
SDSS J1027+1749.
\label{table:host} }
\tablehead{ \colhead{} & \colhead{} & \colhead{} &
\colhead{$r$} & \colhead{$M_r$} & \colhead{$M_{r,c}$} &
\colhead{$u - r$} & \colhead{$R_e$} & \colhead{log$M_{\ast}$} &
\colhead{log$M_{\ast,c}$} & \colhead{$\sigma_{\ast}$} &
\colhead{log$M_{\bullet}$} \\
\colhead{~~~~~~~~~~ID~~~~~~~~~~} & \colhead{SDSS Designation} &
\colhead{$z$} & \colhead{(mag)} & \colhead{(mag)} &
\colhead{(mag)} & \colhead{(mag)} & \colhead{(kpc)} &
\colhead{($M_{\odot}$)} & \colhead{($M_{\odot}$)} &
\colhead{(km s$^{-1}$)} &
\colhead{($M_{\odot}$)}  \\
\colhead{(1)} & \colhead{(2)} & \colhead{(3)} & \colhead{(4)} &
\colhead{(5)} & \colhead{(6)} & \colhead{(7)} & \colhead{(8)} &
\colhead{(9)} & \colhead{(10)} & \colhead{(11)} &
\colhead{(12)} } \startdata
A\dotfill & J102700.40+174900.8\dotfill & 0.0668 &16.73 &
$-20.66$ &
$-22.4$ & 2.48 & 1.1 & $10.4$ & $11.2$& $265\pm6$ & $8.6$ \\
B\dotfill & J102700.55+174900.2\dotfill & 0.0664 &17.00 &
$-20.38$ &
$-23.1$ & 2.46 & 1.6 & $10.1$ & $11.2$& $193\pm10$ & $8.1$ \\
C\dotfill & J102700.38+174902.6\dotfill & 0.0652 &17.03 &
$-20.30$ &
$-24.5$ & 2.51 & 2.4 & $10.2$ & $12.0$& $305\pm12$ & $8.9$ \\
\enddata
\tablecomments{Col. 1: ID of the three nuclei as labeled on
Figure \ref{fig:sdssimg}; Col. 2: SDSS names with J2000
coordinates given in the form of ``hhmmss.ss+ddmmss.s''; Col.
3: redshift measured from stellar continuum fitting; Col. 4:
SDSS $r$-band fiber magnitude; Col. 5: SDSS $r$-band fiber
absolute magnitude; Col. 6: Col. 5 after correction for
internal dust extinction estimated using the Balmer decrement
method; Col. 7: SDSS color from fiber magnitudes; Col. 8:
effective radius from aperture photometry; Col. 9: stellar mass
estimates from population synthesis modeling of the optical
stellar continuum; Col. 10: Col. 9 after correction for
internal dust extinction; Col. 11: stellar velocity dispersion
from model fits of the stellar continuum over the G-band
$\lambda4304$ \angstrom\ region (Figure \ref{fig:dis1d}). The
quoted errors throughout this table are statistical
uncertainties; Col. 12: black hole mass estimates assuming the
$M_{\bullet}$--$\sigma_{\ast}$ relation of \citet{tremaine02}
observed in local inactive galaxies.}
\end{deluxetable*}
%\end{landscape}
%%
%%

%\begin{landscape}
\begin{deluxetable*}{ccccccccccccc}
\tabletypesize{\footnotesize}
%\tabletypesize{\scriptsize}
\tablecolumns{12} \tablewidth{\textwidth}
%\tablewidth{0pc}
%
\tablecaption{Emission-line Measurements of the Three Nuclei of
SDSS J1027+1749.
\label{table:emi} }
\tablehead{ \colhead{} & \colhead{log$L_{{\rm
[O\,\,{\scriptscriptstyle III}]}}$} & \colhead{log$L_{{\rm
[O\,\,{\scriptscriptstyle III}]},c}$} & \colhead{$\sigma_{{\rm
gas}}$} & \colhead{} & \colhead{} & \colhead{} & \colhead{} &
\colhead{} & \colhead{EW$_{{\rm H}\alpha}$} &
\colhead{log $n_e$} &
\colhead{$E(B-V)$} \\
\colhead{~~~ID~~~} & \colhead{($L_{\odot}$)} &
\colhead{($L_{\odot}$)} & \colhead{(km s$^{-1}$)} &
\colhead{${\rm [O\,{\scriptscriptstyle III}]}$/H$\beta$} &
\colhead{${\rm [N\,{\scriptscriptstyle II}]}$/H$\alpha$} &
\colhead{${\rm [O\,{\scriptscriptstyle I}]}$/H$\alpha$} &
\colhead{${\rm [S\,{\scriptscriptstyle II}]}$/H$\alpha$} &
\colhead{H$\alpha$/H$\beta$} & \colhead{(\angstrom )} &
\colhead{(cm$^{-3}$)} &
\colhead{(mag)} \\
\colhead{(1)} & \colhead{(2)} & \colhead{(3)} & \colhead{(4)} &
\colhead{(5)} & \colhead{(6)} & \colhead{(7)} & \colhead{(8)} &
\colhead{(9)} & \colhead{(10)} & \colhead{(11)} &
\colhead{(12)} } \startdata
A\dotfill & $6.35$ & $7.2$ & $248\pm17$ & $1.2\pm0.1$ &
$0.98\pm0.02$ & $0.18\pm0.01$  & $0.61\pm0.02$ & $5.3\pm0.4$ &
$4.5\pm0.1$  &
$-2.0$             & $0.61_{-0.09}^{+0.06}$ \\
B\dotfill & $6.38$ & $7.7$ & $211\pm6$  &$0.52\pm0.03$&
$0.66\pm0.01$ & $0.074\pm0.003$& $0.35\pm0.01$ & $7.2\pm0.2$ &
$19.1\pm0.1$
&0.3$^{+1.3}_{-2.3}$ & $0.93_{-0.03}^{+0.02}$ \\
C\dotfill & $6.42$ & $8.4$ & $301\pm23$ & $3.2\pm0.6$ &
$1.22\pm0.01$ & $0.21\pm0.01$  & $0.81\pm0.02$ & $12.1\pm2.0$&
$6.8\pm0.1$
&1.9$^{+0.2}_{-0.7}$ & $1.5_{-0.3}^{+0.1}$    \\
\enddata
\tablecomments{Col. 1: same as Col. 1 in Table
\ref{table:host}; Col. 2: \OIIIb\ luminosity from Gaussian fits
to the emission lines; Col. 3: Col. 2 after correction for
internal dust extinction estimated using the Balmer decrement
method as listed in Col. 12; Col. 4: gas velocity dispersion
from Gaussian fits of the continuum-subtracted emission lines
(Figure \ref{fig:dis1d}). The quoted 1 $\sigma$ errors
throughout this table are statistical uncertainties; Cols.
5--9: intensity ratios from Gaussian fits of the emission
lines; Col. 10: rest-frame \halpha\ equivalent width; Col. 11:
electron density estimated using the emission-line intensity
ratio \SIIa /\SIIb . Col. 12: color excess estimated from the
emission-line intensity ratio \halpha /\hbeta\ (Col. 9) using
the Balmer decrement method, assuming the intrinsic case B
values of 2.87 for $T = 10^4$ K \citep{osterbrock89} and the
extinction curve of \citet{cardelli89} with $R_{V} = 3.1$.}
\end{deluxetable*}
%\end{landscape}

%%
%%

Tables \ref{table:host} and \ref{table:emi} list our
measurements of the host and emission-line properties of the
three nuclei of SDSS J1027+1749, respectively. The redshifts
measured from stellar continuum fitting show that the
components B and C are offset from A by $\sim450$ and $\sim110$
km s$^{-1}$ in velocity. Their stellar velocity dispersions
$\sigma_{\ast}$ are 200--300 km s$^{-1}$, typical of massive
galaxies at $z\sim0.1$ of similar luminosities and sizes
\citep{bernardi03c}.  To securely separate the three
components, we applied aperture photometry for them
individually with $3''$ diameter (corresponding to 3.8 kpc)
apertures using their SDSS images. They all fall in the red
sequence on the $u-r$, $M_r$ color--magnitude diagram
\citep{baldry04}. We have also performed surface brightness
profile fitting using the \textsc{galfit} package
\citep{peng10}, using stars within the field to model the
point-spread function (PSF). Assuming a five-component model
for the entire galaxy (four S\'{e}rsic profiles for the three
nuclei and the surrounding diffuse component, and an
exponential disk for the tidal feature), we estimate their
$r$-band S\'{e}rsic indices as 1.4, 4.1, and 1.2 for A, B, and
C, respectively. While the fit seems reasonable with a reduced
$\chi^2$ of 1.2, the results are highly uncertain given the
limited resolution ($r$-band PSF FWHM $\sim1.''1$) of the SDSS
images and systematic uncertainties from model degeneracy. {\it
Hubble Space Telescope (HST)} and/or ground-based imaging
assisted with adaptive optics will help better determine their
structural properties. We estimate stellar mass $M_{\ast}$
based on population synthesis modeling of the optical stellar
continuum. The stellar mass ratio of the triple system ${\rm
A:B:C}$ is $2:1:1.3$ before extinction correction and is
$1:1:6$ after. We estimate internal dust extinction from the
emission-line intensity ratio \halpha /\hbeta\ using the Balmer
decrement method \citep{osterbrock89} assuming the intrinsic
case B values of 2.87 for $T = 10^4$ K \citep{osterbrock89} and
the extinction curve of \citet{cardelli89} with $R_{V} = 3.1$.
However, the estimates are highly uncertain due to the
systematic uncertainty in the dust geometry. The three stellar
components are blended in Two Micron All Sky Survey
\citep[2MASS;][]{skrutskie06}. Higher resolution IR imaging is
needed to better constrain their individual stellar masses.

\begin{figure}
  \centering
    \includegraphics[width=82mm]{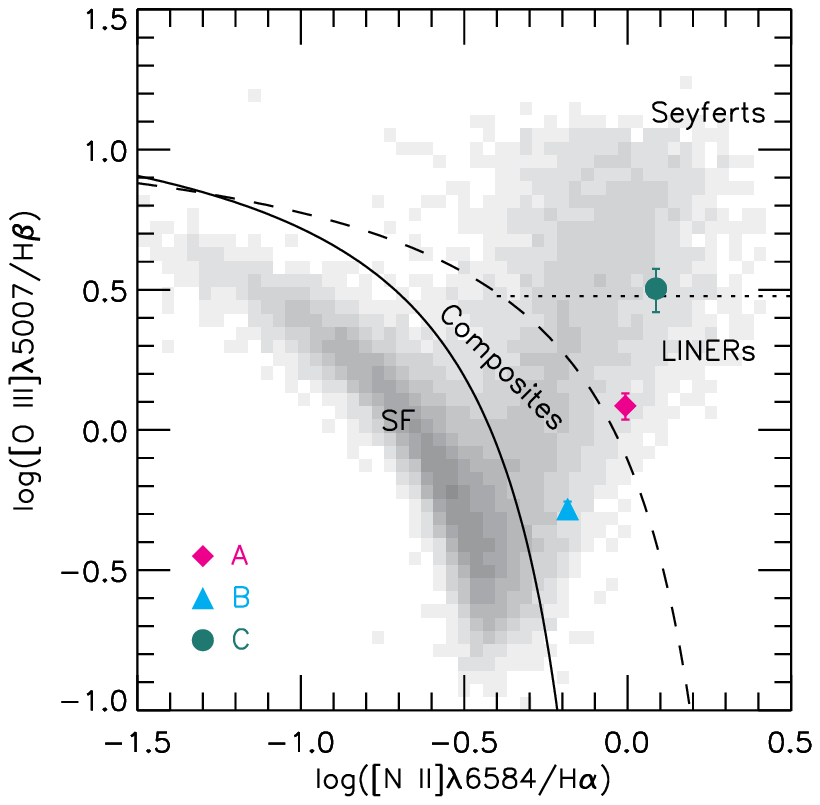}
    \caption{Optical diagnostic emission-line ratios for the three nuclei
    measured from the continuum-subtracted slit spectra (Figure \ref{fig:dis1d}).
    Gray scales indicate number densities of 31,179 emission-line galaxies
    from the SDSS DR4 \citep{kauffmann03}. The solid curve is the empirical
    separation between H {\tiny II} regions and AGNs \citep{kauffmann03}, and the
    dashed curve is the theoretical ``starburst limit'' \citep{kewley06}.
    Pure star-forming (SF) galaxies lie below the solid curve, AGN-dominated
    galaxies lie above the dashed curve, and AGN-H {\tiny II} composites lie
    in between. The dotted curve is the empirical separation between Seyferts
    and LINERs \citep{ho97}.}
    \label{fig:bpt}
\end{figure}

Figure \ref{fig:dis1d} shows that all the three nuclei emit
narrow lines with gas velocity dispersions of $\sigma_{{\rm
gas}}\sim$ 200--300 km s$^{-1}$ (Table \ref{table:emi}),
consistent with $\sigma_{\ast}$ of their associated stellar
components. Their emission line intensity ratios (Table
\ref{table:emi}) are all characteristics of AGNs or AGN-H
{\tiny II} composites.  The masses of the AGNs are estimated to
be $M_{\bullet} \gtrsim 10^8 M_{\odot}$ (Table
\ref{table:host}) from $\sigma_{\ast}$ of their associated
stellar components\footnote{This still holds after correcting
for the fact that $M_{\bullet}$ is likely overestimated by
$\sim0.2$ dex using $\sigma_{\ast}$ measured in AGN mergers
separated by a few kpc \citep{liu11b}.}. Figure \ref{fig:bpt}
shows the BPT diagram \citep{bpt,veilleux87} for the three
nuclei. Based on \NIIb /\halpha\ and \OIIIb /\hbeta , A is
classified as a low-ionization narrow emission region (LINER),
B is a composite, and C is a Seyfert (or a LINER within
uncertainties). A (C) is consistently classified as a LINER
(Seyfert) based on diagnostic ratios \SIIab /\halpha\ or \OI
/\halpha\ and \OIIIb /\hbeta\ according to the criteria of
\citet{kewley06}. While LINERs and composites may be due to
stellar or shock heating rather than AGN excitation
\citep[e.g.,][]{lutz99}, X-ray observations show that
$\sim80$\% of them do harbor AGNs \citep[e.g.,][]{ho08}.
Because the intrinsic AGN luminosities of the three nuclei are
obscured in the optical, we estimate the intrinsic hard X-ray
luminosity $L_{X,{\rm 2-10\,keV}}$ using the \OIIIb\ luminosity
(corrected for internal extinction; Table \ref{table:emi}) as a
surrogate\footnote{Measurements of $L_{X,{\rm
2-10\,keV}}/L_{{\rm [O\,{\scriptscriptstyle III}]}}$ for
optically selected obscured AGNs range from a few to a few
hundred \citep[e.g.,][]{heckman05,panessa06}. Here, we adopt
the mean calibration of \citet{panessa06}, $\log (L_{X,{\rm
2-10\,keV}}/{\rm erg\,s^{-1}}) = 1.22\log (L_{{\rm
[O\,{\scriptscriptstyle III}]}}/{\rm erg\,s^{-1}}) - 7.34$.},
finding $\log (L_{X,{\rm 2-10\,keV}}/{\rm erg\,s^{-1}}) \approx
42.4$, $43.0$, and $43.9$ for A, B, and C, respectively. The
non-detection of the system in the {\it ROSAT} All-Sky Survey
\citep{voges99} is consistent with these estimates. The
intrinsic \OIIIb\ luminosities and \halpha\ equivalent widths
(Table \ref{table:emi}) are typical of $z\sim0.1$ obscured AGNs
\citep[e.g.,][]{kauffmann03,ho08}. The component B has a strong
post-starburst component, as indicated by the strong Balmer
absorption in our slit spectrum. This is also seen in its SDSS
fiber spectrum which covers the higher order Balmer series and
the Balmer break. B is also detected in the FIRST 1.4 GHz
survey with an integrated flux density of $13.57\pm0.14$ mJy
\citep{becker95}. The galaxy is contained in the {\it IRAS}
faint source catalog \citep{moshir92} with a positional match
centered on B, IRAS 10242+1804, with an IR (1--1000 $\mu$m)
luminosity of $L_{{\rm IR}} \sim 10^{11.3} L_{\odot}$ estimated
based on its flux densities at 60 and 100 $\mu$m from {\it
IRAS} using the formulae of \citet{helou88}.

%%%%%%%%%%%%%%%%%%%%%%%%%%%%%%%%%%%%%%%%%%%%%%%%%%%
\section{Discussion}\label{sec:discuss}

Based on dynamical friction timescale
estimates\footnote{Assuming singular isothermal spheres for the
density profiles and circular orbits, the dynamical friction
timescale for a satellite galaxy with velocity dispersion
$\sigma_{{\rm S}}$ inspiraling from radius $r$ in a host galaxy
with velocity dispersion $\sigma_{{\rm H}}$ can be estimated as
$t_{{\rm fric}} = \frac{2.7}{\ln \Lambda}\frac{r}{30\,{\rm
kpc}}\bigg(\frac{\sigma_{{\rm H}}}{200\, {\rm
km\,s^{-1}}}\bigg)^2\bigg(\frac{100\, {\rm
km\,s^{-1}}}{\sigma_{{\rm S}}}\bigg)^3$ Gyr, where $\Lambda$ is
the Coulomb logarithm \citep{binney87}. $\ln \Lambda \sim 2$
for equal mass mergers \citep{dubinski99}. We assume A as the
host, and B and C as satellites, since A is at the apparent
center of the galaxy.  Assuming C as the host instead, which
has the largest $\sigma_{\ast}$, yields similar estimates for
$t_{{\rm fric}}$, $\sim20$ Myr for A and C, and $\sim70$ Myr
for B and C, respectively.}, the stellar components B and C
will merge with A in $t_{{\rm fric}}\sim 40$ and $\sim 8$ Myr,
respectively, corrected for projection assuming random
orientation for $r_p$. The three SMBHs will inspiral with their
individual stellar bulges at first, but will then decay to the
center of the merged galaxy under dynamical friction with the
stellar background in $\sim40$--200 Myr\footnote{Assuming a
singular isothermal sphere for the density distribution of the
host galaxy, the dynamical friction timescale of a BH of mass
$M_{\bullet}$ on a circular orbit of radius $r$ can be
estimated as $t_{{\rm fric}} = \frac{19}{\ln
\Lambda}\bigg(\frac{r}{5\,{\rm kpc}}\bigg)^2\frac{\sigma_{{\rm
H}}}{200\,{\rm km\,s^{-1}}}\frac{10^8\,M_{\odot}}{M_{\bullet}}$
Gyr, where $\ln \Lambda \sim 6$ for typical values
\citep{binney87}. We estimate that $t_{{\rm fric}} \sim 200$
and $40$ Myr for the SMBHs in B and C, to reach the center of
the merger products of B and C with A, respectively.}, given
their mass estimates (Table \ref{table:host}). However, the
actual evolution involving three comparable stellar components
and the associated gas is likely more complicated. High
resolution kinematics observations \citep[e.g.,][]{amram07}
combined with tailored numerical modeling
\citep[e.g.,][]{renaud10} are needed to unravel the merger
history and future of this triple system. Regardless of the
actual sequence of the mergers, the SMBHs may form a
gravitationally interacting triple system, if the orbit of the
third SMBH decays rapidly before the first two SMBHs merge.

The fraction of kpc-scale pairs of AGNs is $f_{{\rm kpc,\,
double}}\gtrsim 3\times10^{-3}$ corrected for SDSS
spectroscopic incompleteness based on the number of AGN pairs
with $r_p<10$ kpc found in our parent AGN sample
(\citealt{liu11a}; see also \citealt{liu10b} and
\citealt{shen10b} for a similar estimate of $f_{{\rm kpc,\,
double}}$ from a complementary approach to identify kpc-scale
AGN pairs, based on the selection of AGNs with double-peaked
narrow emission lines). This is a lower limit due to the
limited imaging resolution of SDSS. Out of the seven kpc-scale
triple AGN candidates, we have obtained slit spectra for five
systems, including SDSS J1027+1749. The nature of the other
four systems seems less clear, due to alternative scenarios
involving star clusters or tidal knots, the results of which
will be presented in a future paper. Our result suggests that
the frequency of kpc-scale triples in optically selected AGNs
at $z\sim0.1$ is $f_{{\rm kpc,\, triple}}\gtrsim 3\times10^{-3}
\times \frac{7}{92} \times \frac{1}{5} \approx 5\times10^{-5}$.
If we assume that AGNs in successive mergers are independent
events with short duty cycles, the expected frequency of
kpc-scale triple AGNs based on the observed frequency of
doubles would be $\sim f_{{\rm kpc,\, double}}^2 \gtrsim
10^{-5}$. Our observed frequency of triples implies a higher
probability than this, which may suggest that AGNs in mergers
are somewhat correlated, although better statistics is needed
to draw any firm conclusion.

While our slit spectra of the three nuclei suggest that they
all host AGNs, optical identification alone is inconclusive.
Alternatively, there could be only two or even just one active
SMBH, ionizing all the three gas components in the merging
system. To discriminate between these scenarios, arguments
based on the ionization parameter and effective size of the
narrow-line region are not very useful
\citep[e.g.,][]{liu10b,shen10b}, because the separation between
the nuclei are not much larger than the individual size of the
gas-emitting regions, and the electron density measurements are
highly uncertain (Table \ref{table:emi}). {\it Chandra} X-ray
observations could help pin down the triple AGN nature of the
system.

%------------------------------------------------------------------------------
\acknowledgments

We thank an anonymous referee for a prompt and helpful report.
Support for the work of X.L. was provided by NASA through
Einstein Postdoctoral Fellowship grant number PF0-110076
awarded by the Chandra X-ray Center, which is operated by the
Smithsonian Astrophysical Observatory for NASA under contract
NAS8-03060. Y.S. acknowledges support from a Clay Postdoctoral
Fellowship through the Smithsonian Astrophysical Observatory.
M.A.S. acknowledges the support of NSF grant AST-0707266.

Funding for the SDSS and SDSS-II has been provided by the
Alfred P. Sloan Foundation, the Participating Institutions, the
National Science Foundation, the U.S. Department of Energy, the
National Aeronautics and Space Administration, the Japanese
Monbukagakusho, the Max Planck Society, and the Higher
Education Funding Council for England. The SDSS Web site is
http://www.sdss.org/.

The SDSS is managed by the Astrophysical Research Consortium
for the Participating Institutions. The Participating
Institutions are the American Museum of Natural History,
Astrophysical Institute Potsdam, University of Basel,
University of Cambridge, Case Western Reserve University,
University of Chicago, Drexel University, Fermilab, the
Institute for Advanced Study, the Japan Participation Group,
Johns Hopkins University, the Joint Institute for Nuclear
Astrophysics, the Kavli Institute for Particle Astrophysics and
Cosmology, the Korean Scientist Group, the Chinese Academy of
Sciences (LAMOST), Los Alamos National Laboratory, the
Max-Planck-Institute for Astronomy (MPIA), the
Max-Planck-Institute for Astrophysics (MPA), New Mexico State
University, Ohio State University, University of Pittsburgh,
University of Portsmouth, Princeton University, the United
States Naval Observatory, and the University of Washington.

{\it Facilities}: Sloan, ARC (DIS)



\bibliography{triple}

\begin{thebibliography}{50}
\expandafter\ifx\csname natexlab\endcsname\relax\def\natexlab#1{#1}\fi

\bibitem[{{Abazajian} {et~al.}(2009){Abazajian},
    {Adelman-McCarthy},
  {Ag{\"u}eros}, {Allam}, {Allende Prieto}, {An}, {Anderson}, {Anderson},
  {Annis}, {Bahcall}, {Bailer-Jones}, {Barentine}, {Bassett}, {Becker},
  {Beers}, {Bell}, {Belokurov}, {Berlind}, {Berman}, {Bernardi}, {Bickerton},
  {Bizyaev}, {Blakeslee}, {Blanton}, {Bochanski}, {Boroski}, {Brewington},
  {Brinchmann}, {Brinkmann}, {Brunner}, {Budav{\'a}ri}, {Carey}, {Carliles},
  {Carr}, {Castander}, {Cinabro}, {Connolly}, {Csabai}, {Cunha}, {Czarapata},
  {Davenport}, {de Haas}, {Dilday}, {Doi}, {Eisenstein}, {Evans}, {Evans},
  {Fan}, {Friedman}, {Frieman}, {Fukugita}, {G{\"a}nsicke}, {Gates},
  {Gillespie}, {Gilmore}, {Gonzalez}, {Gonzalez}, {Grebel}, {Gunn},
  {Gy{\"o}ry}, {Hall}, {Harding}, {Harris}, {Harvanek}, {Hawley}, {Hayes},
  {Heckman}, {Hendry}, {Hennessy}, {Hindsley}, {Hoblitt}, {Hogan}, {Hogg},
  {Holtzman}, {Hyde}, {Ichikawa}, {Ichikawa}, {Im}, {Ivezi{\'c}}, {Jester},
  {Jiang}, {Johnson}, {Jorgensen}, {Juri{\'c}}, {Kent}, {Kessler}, {Kleinman},
  {Knapp}, {Konishi}, {Kron}, {Krzesinski}, {Kuropatkin}, {Lampeitl},
  {Lebedeva}, {Lee}, {Lee}, {Leger}, {L{\'e}pine}, {Li}, {Lima}, {Lin}, {Long},
  {Loomis}, {Loveday}, {Lupton}, {Magnier}, {Malanushenko}, {Malanushenko},
  {Mandelbaum}, {Margon}, {Marriner}, {Mart{\'{\i}}nez-Delgado}, {Matsubara},
  {McGehee}, {McKay}, {Meiksin}, {Morrison}, {Mullally}, {Munn}, {Murphy},
  {Nash}, {Nebot}, {Neilsen}, {Newberg}, {Newman}, {Nichol}, {Nicinski},
  {Nieto-Santisteban}, {Nitta}, {Okamura}, {Oravetz}, {Ostriker}, {Owen},
  {Padmanabhan}, {Pan}, {Park}, {Pauls}, {Peoples}, {Percival}, {Pier}, {Pope},
  {Pourbaix}, {Price}, {Purger}, {Quinn}, {Raddick}, {Fiorentin}, {Richards},
  {Richmond}, {Riess}, {Rix}, {Rockosi}, {Sako}, {Schlegel}, {Schneider},
  {Scholz}, {Schreiber}, {Schwope}, {Seljak}, {Sesar}, {Sheldon}, {Shimasaku},
  {Sibley}, {Simmons}, {Sivarani}, {Smith}, {Smith}, {Smol{\v c}i{\'c}},
  {Snedden}, {Stebbins}, {Steinmetz}, {Stoughton}, {Strauss}, {Subba Rao},
  {Suto}, {Szalay}, {Szapudi}, {Szkody}, {Tanaka}, {Tegmark}, {Teodoro},
  {Thakar}, {Tremonti}, {Tucker}, {Uomoto}, {Vanden Berk}, {Vandenberg},
  {Vidrih}, {Vogeley}, {Voges}, {Vogt}, {Wadadekar}, {Watters}, {Weinberg},
  {West}, {White}, {Wilhite}, {Wonders}, {Yanny}, {Yocum}, {York}, {Zehavi},
  {Zibetti}, \& {Zucker}}]{SDSSDR7}
{Abazajian}, K.~N., {et~al.} 2009, \apjs, 182, 543

\bibitem[{{Amaro-Seoane} {et~al.}(2010){Amaro-Seoane},
    {Sesana}, {Hoffman},
  {Benacquista}, {Eichhorn}, {Makino}, \& {Spurzem}}]{amaro10}
{Amaro-Seoane}, P., {Sesana}, A., {Hoffman}, L., {Benacquista},
M., {Eichhorn},
  C., {Makino}, J., \& {Spurzem}, R. 2010, \mnras, 402, 2308

\bibitem[{{Amram} {et~al.}(2007){Amram}, {Mendes de Oliveira},
    {Plana},
  {Balkowski}, \& {Hernandez}}]{amram07}
{Amram}, P., {Mendes de Oliveira}, C., {Plana}, H.,
{Balkowski}, C., \&
  {Hernandez}, O. 2007, \aap, 471, 753

\bibitem[{{Baldry} {et~al.}(2004){Baldry}, {Glazebrook},
    {Brinkmann},
  {Ivezi{\'c}}, {Lupton}, {Nichol}, \& {Szalay}}]{baldry04}
{Baldry}, I.~K., {Glazebrook}, K., {Brinkmann}, J.,
{Ivezi{\'c}}, {\v Z}.,
  {Lupton}, R.~H., {Nichol}, R.~C., \& {Szalay}, A.~S. 2004, \apj, 600, 681

\bibitem[{{Baldwin} {et~al.}(1981){Baldwin}, {Phillips}, \&
    {Terlevich}}]{bpt} {Baldwin}, J.~A., {Phillips}, M.~M., \&
    {Terlevich}, R. 1981, \pasp, 93, 5

\bibitem[{{Barth} {et~al.}(2008){Barth}, {Bentz}, {Greene}, \&
    {Ho}}]{barth08} {Barth}, A.~J., {Bentz}, M.~C., {Greene},
    J.~E., \& {Ho}, L.~C. 2008, \apjl,
  683, L119

\bibitem[{{Becker} {et~al.}(1995){Becker}, {White}, \&
    {Helfand}}]{becker95} {Becker}, R.~H., {White}, R.~L., \&
    {Helfand}, D.~J. 1995, \apj, 450, 559

\bibitem[{{Begelman} {et~al.}(1980){Begelman}, {Blandford}, \&
  {Rees}}]{begelman80}
{Begelman}, M.~C., {Blandford}, R.~D., \& {Rees}, M.~J. 1980,
\nat, 287, 307

\bibitem[{{Bernardi} {et~al.}(2003){Bernardi}, {Sheth},
    {Annis}, {Burles},
  {Eisenstein}, {Finkbeiner}, {Hogg}, {Lupton}, {Schlegel}, {SubbaRao},
  {Bahcall}, {Blakeslee}, {Brinkmann}, {Castander}, {Connolly}, {Csabai},
  {Doi}, {Fukugita}, {Frieman}, {Heckman}, {Hennessy}, {Ivezi{\'c}}, {Knapp},
  {Lamb}, {McKay}, {Munn}, {Nichol}, {Okamura}, {Schneider}, {Thakar}, \&
  {York}}]{bernardi03c}
{Bernardi}, M., {et~al.} 2003, \aj, 125, 1866

\bibitem[{{Binney} \& {Tremaine}(1987)}]{binney87} {Binney},
    J., \& {Tremaine}, S. 1987, {Galactic dynamics}, ed.
    {Binney, J.~\&
  Tremaine, S.}

\bibitem[{{Blaes} {et~al.}(2002){Blaes}, {Lee}, \&
    {Socrates}}]{blaes02} {Blaes}, O., {Lee}, M.~H., \&
    {Socrates}, A. 2002, \apj, 578, 775

\bibitem[{{Bruzual} \& {Charlot}(2003)}]{bc03} {Bruzual}, G.,
    \& {Charlot}, S. 2003, \mnras, 344, 1000

\bibitem[{{Cardelli} {et~al.}(1989){Cardelli}, {Clayton}, \&
  {Mathis}}]{cardelli89}
{Cardelli}, J.~A., {Clayton}, G.~C., \& {Mathis}, J.~S. 1989,
\apj, 345, 245

\bibitem[{{Djorgovski} {et~al.}(2007){Djorgovski}, {Courbin},
    {Meylan},
  {Sluse}, {Thompson}, {Mahabal}, \& {Glikman}}]{djorgovski07}
{Djorgovski}, S.~G., {Courbin}, F., {Meylan}, G., {Sluse}, D.,
{Thompson}, D.,
  {Mahabal}, A., \& {Glikman}, E. 2007, \apjl, 662, L1

\bibitem[{{Dubinski} {et~al.}(1999){Dubinski}, {Mihos}, \&
  {Hernquist}}]{dubinski99}
{Dubinski}, J., {Mihos}, J.~C., \& {Hernquist}, L. 1999, \apj,
526, 607

\bibitem[{{Greene} \& {Ho}(2006)}]{greene06a} {Greene}, J.~E.,
    \& {Ho}, L.~C. 2006, \apj, 641, 117

\bibitem[{{Heckman} {et~al.}(2005){Heckman}, {Ptak},
    {Hornschemeier}, \&
  {Kauffmann}}]{heckman05}
{Heckman}, T.~M., {Ptak}, A., {Hornschemeier}, A., \&
{Kauffmann}, G. 2005,
  \apj, 634, 161

\bibitem[{{Helou} {et~al.}(1988){Helou}, {Khan}, {Malek}, \&
  {Boehmer}}]{helou88}
{Helou}, G., {Khan}, I.~R., {Malek}, L., \& {Boehmer}, L. 1988,
\apjs, 68, 151

\bibitem[{{Hernquist}(1989)}]{hernquist89} {Hernquist}, L.
    1989, \nat, 340, 687

\bibitem[{{Ho}(2008)}]{ho08} {Ho}, L.~C. 2008, \araa, 46, 475

\bibitem[{{Ho} {et~al.}(1997){Ho}, {Filippenko}, \&
    {Sargent}}]{ho97} {Ho}, L.~C., {Filippenko}, A.~V., \&
    {Sargent}, W.~L.~W. 1997, \apjs, 112, 315

\bibitem[{{Ho} {et~al.}(2009){Ho}, {Greene}, {Filippenko}, \&
    {Sargent}}]{ho09} {Ho}, L.~C., {Greene}, J.~E.,
    {Filippenko}, A.~V., \& {Sargent}, W.~L.~W. 2009,
  \apjs, 183, 1

\bibitem[{{Hoffman} \& {Loeb}(2007)}]{hoffman07} {Hoffman}, L.,
    \& {Loeb}, A. 2007, \mnras, 377, 957

\bibitem[{{Iwasawa} {et~al.}(2006){Iwasawa}, {Funato}, \&
    {Makino}}]{iwasawa06} {Iwasawa}, M., {Funato}, Y., \&
    {Makino}, J. 2006, \apj, 651, 1059

\bibitem[{{Kauffmann} {et~al.}(2003){Kauffmann}, {Heckman},
    {Tremonti},
  {Brinchmann}, {Charlot}, {White}, {Ridgway}, {Brinkmann}, {Fukugita}, {Hall},
  {Ivezi{\'c}}, {Richards}, \& {Schneider}}]{kauffmann03}
{Kauffmann}, G., {et~al.} 2003, \mnras, 346, 1055

\bibitem[{{Kewley} {et~al.}(2006){Kewley}, {Groves},
    {Kauffmann}, \&
  {Heckman}}]{kewley06}
{Kewley}, L.~J., {Groves}, B., {Kauffmann}, G., \& {Heckman},
T. 2006, \mnras,
  372, 961

\bibitem[{{Kormendy} \& {Richstone}(1995)}]{kormendy95}
    {Kormendy}, J., \& {Richstone}, D. 1995, \araa, 33, 581

\bibitem[{{Liu} {et~al.}(2010){Liu}, {Greene}, {Shen}, \&
    {Strauss}}]{liu10b} {Liu}, X., {Greene}, J.~E., {Shen}, Y.,
    \& {Strauss}, M.~A. 2010, \apjl, 715,
  L30

\bibitem[{{Liu} {et~al.}(2011{\natexlab{a}}){Liu}, {Shen}, \&
  {Strauss}}]{liu11b}
{Liu}, X., {Shen}, Y., \& {Strauss}, M.~A. 2011{\natexlab{a}},
arXiv:1104.0951

\bibitem[{{Liu} {et~al.}(2011{\natexlab{b}}){Liu}, {Shen},
    {Strauss}, \&
  {Hao}}]{liu11a}
{Liu}, X., {Shen}, Y., {Strauss}, M.~A., \& {Hao}, L.
2011{\natexlab{b}}, arXiv:1104.0950

\bibitem[{{Liu} {et~al.}(2009){Liu}, {Zakamska}, {Greene},
    {Strauss}, {Krolik},
  \& {Heckman}}]{liu09}
{Liu}, X., {Zakamska}, N.~L., {Greene}, J.~E., {Strauss},
M.~A., {Krolik},
  J.~H., \& {Heckman}, T.~M. 2009, \apj, 702, 1098

\bibitem[{{Lousto} \& {Zlochower}(2008)}]{lousto08} {Lousto},
    C.~O., \& {Zlochower}, Y. 2008, \prd, 77, 024034

\bibitem[{{Lutz} {et~al.}(1999){Lutz}, {Veilleux}, \&
    {Genzel}}]{lutz99} {Lutz}, D., {Veilleux}, S., \& {Genzel},
    R. 1999, \apjl, 517, L13

\bibitem[{{Merritt}(2006)}]{merritt06} {Merritt}, D. 2006,
    Reports on Progress in Physics, 69, 2513

\bibitem[{{Milosavljevi{\'c}} \&
    {Merritt}(2001)}]{milosavljevic01} {Milosavljevi{\'c}}, M.,
    \& {Merritt}, D. 2001, \apj, 563, 34

\bibitem[{{Moshir} {et~al.}(1992){Moshir}, {Kopman}, \&
    {Conrow}}]{moshir92} {Moshir}, M., {Kopman}, G., \&
    {Conrow}, T.~A.~O. 1992, {IRAS Faint Source
  Survey, Explanatory supplement version 2}, ed. {Moshir, M., Kopman, G., \&
  Conrow, T.~A.~O.}

\bibitem[{{Osterbrock}(1989)}]{osterbrock89} {Osterbrock},
    D.~E. 1989, {Astrophysics of gaseous nebulae and active
    galactic
  nuclei}, ed. {Osterbrock, D.~E.}

\bibitem[{{Panessa} {et~al.}(2006){Panessa}, {Bassani},
    {Cappi}, {Dadina},
  {Barcons}, {Carrera}, {Ho}, \& {Iwasawa}}]{panessa06}
{Panessa}, F., {Bassani}, L., {Cappi}, M., {Dadina}, M.,
{Barcons}, X.,
  {Carrera}, F.~J., {Ho}, L.~C., \& {Iwasawa}, K. 2006, \aap, 455, 173

\bibitem[{{Peng} {et~al.}(2010){Peng}, {Ho}, {Impey}, \&
    {Rix}}]{peng10} {Peng}, C.~Y., {Ho}, L.~C., {Impey}, C.~D.,
    \& {Rix}, H. 2010, \aj, 139, 2097

\bibitem[{{Renaud} {et~al.}(2010){Renaud}, {Appleton}, \&
    {Xu}}]{renaud10} {Renaud}, F., {Appleton}, P.~N., \& {Xu},
    C.~K. 2010, \apj, 724, 80

\bibitem[{{Shen} {et~al.}(2011){Shen}, {Liu}, {Greene}, \& {Strauss}}]{shen10b}
{Shen}, Y., {Liu}, X., {Greene}, J.~E., \& {Strauss}, M.~A. 2011, \apj, 735, 48

\bibitem[{{Skrutskie} {et~al.}(2006){Skrutskie}, {Cutri},
    {Stiening},
  {Weinberg}, {Schneider}, {Carpenter}, {Beichman}, {Capps}, {Chester},
  {Elias}, {Huchra}, {Liebert}, {Lonsdale}, {Monet}, {Price}, {Seitzer},
  {Jarrett}, {Kirkpatrick}, {Gizis}, {Howard}, {Evans}, {Fowler}, {Fullmer},
  {Hurt}, {Light}, {Kopan}, {Marsh}, {McCallon}, {Tam}, {Van Dyk}, \&
  {Wheelock}}]{skrutskie06}
{Skrutskie}, M.~F., {et~al.} 2006, \aj, 131, 1163

\bibitem[{{Tody}(1986)}]{tody86} {Tody}, D. 1986, in Society of
    Photo-Optical Instrumentation Engineers (SPIE)
  Conference Series, ed. {D.~L.~Crawford}, Vol. 627, 733

\bibitem[{{Toomre} \& {Toomre}(1972)}]{toomre72} {Toomre}, A.,
    \& {Toomre}, J. 1972, \apj, 178, 623

\bibitem[{{Tremaine} {et~al.}(2002){Tremaine}, {Gebhardt},
    {Bender}, {Bower},
  {Dressler}, {Faber}, {Filippenko}, {Green}, {Grillmair}, {Ho}, {Kormendy},
  {Lauer}, {Magorrian}, {Pinkney}, \& {Richstone}}]{tremaine02}
{Tremaine}, S., {et~al.} 2002, \apj, 574, 740

\bibitem[{{Valtonen}(1996)}]{valtonen96} {Valtonen}, M.~J.
    1996, \mnras, 278, 186

\bibitem[{{Veilleux} \& {Osterbrock}(1987)}]{veilleux87}
    {Veilleux}, S., \& {Osterbrock}, D.~E. 1987, \apjs, 63, 295

\bibitem[{{Voges} {et~al.}(1999){Voges}, {Aschenbach},
    {Boller},
  {Br{\"a}uninger}, {Briel}, {Burkert}, {Dennerl}, {Englhauser}, {Gruber},
  {Haberl}, {Hartner}, {Hasinger}, {K{\"u}rster}, {Pfeffermann}, {Pietsch},
  {Predehl}, {Rosso}, {Schmitt}, {Tr{\"u}mper}, \& {Zimmermann}}]{voges99}
{Voges}, W., {et~al.} 1999, \aap, 349, 389

\bibitem[{{York} {et~al.}(2000){York}, {Adelman}, {Anderson},
    {Anderson},
  {Annis}, {Bahcall}, {Bakken}, {Barkhouser}, {Bastian}, {Berman}, {Boroski},
  {Bracker}, {Briegel}, {Briggs}, {Brinkmann}, {Brunner}, {Burles}, {Carey},
  {Carr}, {Castander}, {Chen}, {Colestock}, {Connolly}, {Crocker}, {Csabai},
  {Czarapata}, {Davis}, {Doi}, {Dombeck}, {Eisenstein}, {Ellman}, {Elms},
  {Evans}, {Fan}, {Federwitz}, {Fiscelli}, {Friedman}, {Frieman}, {Fukugita},
  {Gillespie}, {Gunn}, {Gurbani}, {de Haas}, {Haldeman}, {Harris}, {Hayes},
  {Heckman}, {Hennessy}, {Hindsley}, {Holm}, {Holmgren}, {Huang}, {Hull},
  {Husby}, {Ichikawa}, {Ichikawa}, {Ivezi{\'c}}, {Kent}, {Kim}, {Kinney},
  {Klaene}, {Kleinman}, {Kleinman}, {Knapp}, {Korienek}, {Kron}, {Kunszt},
  {Lamb}, {Lee}, {Leger}, {Limmongkol}, {Lindenmeyer}, {Long}, {Loomis},
  {Loveday}, {Lucinio}, {Lupton}, {MacKinnon}, {Mannery}, {Mantsch}, {Margon},
  {McGehee}, {McKay}, {Meiksin}, {Merelli}, {Monet}, {Munn}, {Narayanan},
  {Nash}, {Neilsen}, {Neswold}, {Newberg}, {Nichol}, {Nicinski}, {Nonino},
  {Okada}, {Okamura}, {Ostriker}, {Owen}, {Pauls}, {Peoples}, {Peterson},
  {Petravick}, {Pier}, {Pope}, {Pordes}, {Prosapio}, {Rechenmacher}, {Quinn},
  {Richards}, {Richmond}, {Rivetta}, {Rockosi}, {Ruthmansdorfer}, {Sandford},
  {Schlegel}, {Schneider}, {Sekiguchi}, {Sergey}, {Shimasaku}, {Siegmund},
  {Smee}, {Smith}, {Snedden}, {Stone}, {Stoughton}, {Strauss}, {Stubbs},
  {SubbaRao}, {Szalay}, {Szapudi}, {Szokoly}, {Thakar}, {Tremonti}, {Tucker},
  {Uomoto}, {Vanden Berk}, {Vogeley}, {Waddell}, {Wang}, {Watanabe},
  {Weinberg}, {Yanny}, \& {Yasuda}}]{york00}
{York}, D.~G., {et~al.} 2000, \aj, 120, 1579

\bibitem[{{Yu}(2002)}]{yu02} {Yu}, Q. 2002, \mnras, 331, 935


\end{thebibliography}

\end{document}